\documentclass[twoside,english]{IEEEtran}
\usepackage[T1]{fontenc}
\usepackage{textcomp}
\usepackage[latin9]{inputenc}
\usepackage{geometry}
\usepackage{color}
\usepackage{mathtools}
\usepackage{amsmath}
\usepackage{amssymb}
\usepackage{graphicx}

\makeatletter

\newcommand{\lyxmathsym}[1]{\ifmmode\begingroup\def\b@ld{bold}
  \text{\ifx\math@version\b@ld\bfseries\fi#1}\endgroup\else#1\fi}

\providecommand{\tabularnewline}{\\}

\usepackage{cite}
\makeatother

\usepackage{babel}
\begin{document}
\title{\textcolor{black}{Koopman-Based Model Predictive Control of Functional
Electrical Stimulation for Ankle Dorsiflexion and Plantarflexion Assistance}}
\author{\textcolor{black}{Mayank Singh$^{*}$, Noor Hakam, Trisha M. Kesar$^{\dagger}$,
Nitin Sharma{*}\thanks{\textcolor{black}{M. Singh is with the Department of Electrical Engineering,
North Carolina State University-Raleigh. (Email: msingh25@ncsu.edu).
$^{\dagger}$Trisha Kesar is with the Division of Physical Therapy
Department of Rehabilitation Medicine Emory University School of Medicine,
Atlanta, GA-30322. Nitin Sharma is with the UNC/NC State Joint Department
of Biomedical Engineering, NC State University - Raleigh, NC 27606
USA (e-mail: nsharm23@ncsu.edu).}}\thanks{\textcolor{black}{{*}Corresponding author: Nitin Sharma. This work
was funded by in part by NSF CAREER Award \#2002261, NSF Award \#2124017,
and NIH R21HD116484.}}}}

\maketitle
\textbf{\textcolor{black}{Functional Electrical Stimulation (FES)
can be an effective tool to augment paretic muscle function and restore
normal ankle function. Our approach incorporates a real-time, data-driven
Model Predictive Control (MPC) scheme, built upon a Koopman operator
theory (KOT) framework. This framework adeptly captures the complex
nonlinear dynamics of ankle motion in a linearized form, enabling
application of linear control approaches for highly-nonlinear FES-actuated
dynamics. Utilizing inertial measurement units (IMUs), our method
accurately predicts the FES-induced ankle movements, while accounting
for nonlinear muscle actuation dynamics, including the muscle activation
for both plantarflexors, and dorsiflexors (Tibialis Anterior (TA)).
The linear prediction model derived through KOT allowed us to formulate
the MPC problem with linear state space dynamics, enhancing the real-time
feasibility, precision and adaptability of the FES driven control.
The effectiveness and applicability of our approach have been demonstrated
through comprehensive simulations and experimental trials, including
three participants with no disability and a participant with Multiple
Sclerosis. Our findings highlight the potential of a KOT-based MPC
approach for FES based gait assistance that offers effective and personalized
assistance for individuals with gait impairment conditions.}}

\section{\textcolor{black}{Introduction}}

\textcolor{black}{Neurological conditions such as stroke, spinal cord
injury (SCI), cerebral palsy, and multiple sclerosis (MS) often impair
ankle function, necessitating specialized rehabilitation interventions.
Functional Electrical Stimulation (FES) can restore ankle function
by eliciting artificial muscle contractions in paralyzed plantarflexor
and dorsiflexor muscles through the application of non-invasive electrical
stimulation, and thereby facilitating improved joint function \cite{york2019survey}.}

\textcolor{black}{To effectuate a natural and efficient walking pattern,
FES-based gait assistance requires accurately timing and modulating
electrical stimulation to the Gastrocnemius (GAS) muscle for plantarflexion
during the push-off phase, and to Tibialis Anterior (TA) for dorsiflexion
during the swing phase of the gait cycle. The effectiveness of FES
in gait rehabilitation has been demonstrated through various control
strategies for ankle rehabilitation \cite{hayashibe2011dual,kesar2009functional,luo2020review}.
A comprehensive review for gait assistance using FES based control
methods can be found in \cite{gil2020advances}.}

\textcolor{black}{Among contemporary control methodologies, Iterative
Learning Control (ILC) has been extensively researched for FES-based
joint functionality improvement \cite{seel2016iterative}. ILC schemes
are often model-free or rely on linear time-invariant dynamics, simplifying
implementation. For instance, Seel et. al. \cite{seel2016iterative}
used ILC with six inertial sensors to estimate ankle angles in post-stroke
patients, achieving rapid convergence but introducing discontinuities
by resetting control inputs after each gait cycle and requiring substantial
sensor data. Freeman et. al. \cite{page2020point} improved the ILC
design by developing a continuous repetitive control scheme, eliminating
reinitialization, reducing computational burden, and enhancing trajectory
tracking. Similarly, Jiang et. al. \cite{jiang2020iterative} designed
a framework for dorsiflexion assistance using dual parameters to reduce
stimulation intensity and mitigate muscle fatigue. Müller et. al.
\cite{muller2020adaptive} extended ILC to assist both knee and ankle
motion, showing adaptability for individual stimulation patterns but
requiring refinement to address sensitivity to knee angle resets.}

\textcolor{black}{However, ILC-based FES designs, including those
in \cite{jiang2020iterative} and \cite{page2020point}, typically
assume linear dynamics and overlook disturbances or model uncertainties,
limiting their robustness. Moreover, measurements derived from physiological
sensors induce additional nonlinearities which cannot be easily integrated
in the ILC framework. Nonlinear control of FES for ankle control was
demonstrated by Zhang et. al. \cite{zhang2022ultrasound}, where ultrasound-derived
muscle activation was integrated into a nonlinear model for drop foot
correction. However, their state feedback-based adaptive impedance
controller lacked optimization, risking overstimulation and rapid
muscle fatigue due to unaddressed FES input constraints.}

\textcolor{black}{Moreover, results on FES based gait assistance/improvement
presented in \cite{seel2016iterative,page2020point,jiang2020iterative,muller2020adaptive,zhang2022ultrasound}
have primarily focused on drop foot correction. In these results,
FES mainly targets the TA muscle during the swing phase, foregoing
stimulation of plantarflexors, which are critical for push-off \cite{awad2020central}.
Efficacy of FES stimuli to plantarflexor muscles has been shown for
correcting post-stroke gait deficits \cite{kesar2009functional} and
improving walking after SCI \cite{bajd1994significance}. It was noted
in \cite{kesar2009functional,awad2020central}, that applying stimulation
to both plantarflexors and dorsiflexors results in improved gait that
is closer to the normal gait cycle in chronic stroke survivors. Despite
the evidence on the significance of FES-elicited plantarflexion, closed-loop
control of both FES-evoked plantarflexion and dorsiflexion remains
unexplored.}

\textcolor{black}{In this paper, we present a novel Koopman Model
Predictive Control (KMPC) framework for Functional Electrical Stimulation
(FES)-based gait assistance, applying optimally designed stimulation
signals to both plantarflexor and dorsiflexor muscles throughout the
entire gait cycle. By employing Koopman Operator Theory (KOT), we
captures the system\textquoteright s nonlinearities through the linear
evolution of lifted observable functions of the states, facilitating
the application of linear control techniques for the MPC framework.
We derive a linear representation of the inherently nonlinear ankle
dynamics, enabling real-time prediction and control of the full gait
cycle. The data-driven operator converts the nonlinear ankle motion
dynamics into linear dynamics, which eases MPC formulation and real-time
implementation \cite{proctor2018generalizing,korda2020optimal,korda2018linear}.
The KMPC formulation performs dynamic optimization in real time, utilizing
Koopman-based FES actuated ankle model to predict future system behavior
and solve the moving horizon optimization problem to design optimal
FES stimulation input.}

\textcolor{black}{For ankle assistance control, \cite{benoussaad2013nonlinear}
used an MPC to design optimal muscle excitation for the TA muscle
for adequate toe/foot clearance. While constraints on ankle and control
inputs were considered, a formal closed-loop stability and control
feasibility analysis were missing. As the effectiveness of MPC depends
on model accuracy, necessitating extensive system identification,
especially for complex neuromuscular dynamics. Addressing nonlinear
dynamics and constraints can increase computational demands, posing
challenges for real-time implementation. To mitigate these issues,
\cite{singh2023data} used a Koopman-based data-driven MPC control
to calculate optimal FES stimulation for the TA muscle to correct
drop foot during the swing phase to avoid toe drag. In this paper,
we extend the data-driven MPC to design optimal FES stimulation for
ankle assistance for both plantarflexors and dorsiflexors to provide
assistance during complete gait cycle. To the best of our knowledge,
this is the first implementation of an FES-based optimal control strategy
for the entire gait cycle in real time.}

\textcolor{black}{The paper is organized as follows -- \ref{sec:Ankle-joint-dorsiflexion}
describes the ankle dorsiflexion and plantarflexion motion dynamics
actuated under FES. \ref{sec:Koopman-Based-Model-Predictive} discusses
the overview of the Koopman-based data-driven model of ankle dynamics
and the subsequent formulation of the MPC-based control synthesis
problem in Section \ref{sec:Data-driven-model-predictive}. Experimental
setup, simulation, and experiment results are presented in \ref{sec:Experimental-Results}.
The results are discussed in \ref{sec:Discussion} with with some
discussion on limitations and scope for future directions. Finally,
the paper concludes in \ref{sec:Conclusion}.}

\section{\textcolor{black}{Ankle joint gait dynamics\protect\label{sec:Ankle-joint-dorsiflexion}}}

\textcolor{black}{During a gait cycle, the ankle movement is modeled
as continuous dynamics within swing and stance phases with a discrete
transition event between the two phases. Therefore, the ankle dynamics,
modeled as a switched system to accommodate for the transition, is
given as
\begin{eqnarray}
\begin{bmatrix}J^{P}\ddot{\theta}+f_{J}^{P}(\theta,\dot{\theta})\\
J^{D}\ddot{\theta}+f_{J}^{D}(\theta,\dot{\theta})
\end{bmatrix} & = & \begin{cases}
\tau^{P} & t\in t_{st}\\
\tau^{D}, & t\in t_{sw}
\end{cases}\label{eq:neuromuscular=000020dynamics}
\end{eqnarray}
The net torque about the ankle is defined as $\tau^{\zeta}=g_{J}^{\zeta}(\theta,\dot{\theta}u^{\zeta})\in\mathbb{R}$,
where $\zeta=P,D$ represent the ankle dynamics driven by ankle plantarflexors
during the stance phase and ankle dorsiflexors during the swing phase,
respectively. The net torque terms, $g_{J}^{\zeta}(\theta,\dot{\theta}u^{\zeta})$,
include the torque-angle and torque-angular velocity terms, and and
$u^{\zeta}\in\mathbb{R}$, which is the FES modulated parameter (current,
pulse width, or frequency) applied on the GAS, and TA muscles \cite{kirsch2017nonlinear}.
The stance and swing phases are timed as $t_{st}\coloneqq[t_{start},t_{stance}]$
and $t_{sw}\coloneqq[t_{swing},t_{end}]$, respectively. $J^{\zeta}\in\mathbb{R}^{+}$
is the unknown inertia term of the foot along the dorsiflexion and
plantarflexion axis of rotation, and $\theta(t)$, $\dot{\theta}(t)$,
and $\ddot{\theta}(t)\in\mathbb{R}$ denote the angular position,
angular velocity, and angular acceleration, respectively. $f_{J}^{\zeta}(\theta,\dot{\theta})$
in (\ref{eq:neuromuscular=000020dynamics}) is composed of the musculoskeletal
viscosity torque term, musculoskeletal elasticity, and the gravitational
term. The explicit definitions of the functions can be obtained from
\cite{kirsch2017nonlinear}.}

\textcolor{black}{For each phase $t_{st}$ and $t_{sw}$, we can rewrite
the system dynamics in (\ref{eq:neuromuscular=000020dynamics}), by
selecting $\theta_{1}=\theta$ and $\theta_{2}=\dot{\theta}$.The
equivalent state space representation for can be formulated as
\begin{equation}
\dot{x}_{a}=\begin{cases}
f_{a}^{P}(x_{a})+g_{a}^{P}(x_{a},u^{P}) & \text{\ensuremath{\forall\ }}t\in t_{st}\\
f_{a}^{D}(x_{a})+g_{a}^{D}(x_{a},u^{D}) & \text{\ensuremath{\forall} }t\in t_{sw}
\end{cases}\label{eq:generalized=000020state=000020space}
\end{equation}
where $\dot{x}_{a}=\left[\begin{array}{cc}
\dot{\theta}_{1} & \dot{\theta}_{2}\end{array}\right]^{T}$, $f_{a}^{\zeta}(x_{a})\in\mathbb{R}^{2}$ are the system dynamics,
and $g_{a}^{\zeta}(x_{a},u^{\zeta})\in\mathbb{R}^{2}$ are the actuation
dynamics.}

\textcolor{black}{We can now set up the optimal tracking problem by
defining a tracking error $e(t)\in\mathbb{R}^{2},$ which is defined
as 
\begin{equation}
e=x_{a}-x_{d},\label{eq:Error}
\end{equation}
where $x_{d}\in\mathbb{R}^{2}$ is a bounded desired trajectory for
the desired position and velocity. It is assumed that}\textbf{\textcolor{black}{{}
}}\textcolor{black}{$x_{d}$ and its first derivative, $\dot{x}_{d}=h_{d}(x_{d})\in\mathbb{R}^{2}$,
are bounded.}

\textcolor{black}{By defining an augmented state as $x=\left[\begin{array}{cc}
e^{T} & x_{d}^{T}\end{array}\right]^{T}\in\mathbb{R}^{3},$ the system dynamics can be written as $\dot{x}=f^{\zeta}(x)+g^{\zeta}(x,u^{\zeta}),$
where the system matrices $f^{\zeta}(x)$ and $g^{\zeta}(x,u^{\zeta})$
matrices become $f^{\zeta}(x)=\left[\begin{array}{c}
f^{\zeta}(e+x_{d})-h_{d}(x_{d})\\
h_{d}(x_{d})
\end{array}\right];\:g^{\zeta}(x,u^{\zeta})=\left[\begin{array}{c}
g^{\zeta}(e+x_{d})\\
0
\end{array}\right]$. }

\textcolor{black}{Using zero order hold approximation the continuous-time
system above can be discretized and described as
\begin{align}
x_{k+1}=\begin{cases}
f^{P}(x_{k})+g^{P}(x_{k},u_{k}^{P}) & \text{\ensuremath{\forall\ }}t\in t_{st}\\
f^{D}(x_{k})+g^{D}(x_{k},u_{k}^{D}) & \text{\ensuremath{\forall\ }}t\in t_{sw}
\end{cases}\label{eq:dtsys}
\end{align}
We can define an indicator function $\sigma_{k}$ based on the gait
phase time intervals for the stance and swing phases as
\begin{equation}
\sigma_{k}=\begin{cases}
0 & \text{\ensuremath{\forall\ }}t\in t_{st}\\
1 & \text{\ensuremath{\forall\ }}t\in t_{sw}
\end{cases},\label{eq:indicatorfcn}
\end{equation}
}

\textcolor{black}{where the phase indicator $\sigma_{k}$ takes the
value $0$ for the stance phase and $1$for the swing phase. Upon
incorporating the phase indicator the complete ankle motion dynamics
during a gait cycle can then be described as
\begin{eqnarray}
x_{k+1} & = & (1-\sigma_{k})\left(f^{P}(x_{k})+g^{P}(x_{k},u_{k}^{P})\right)\nonumber \\
 &  & +\sigma_{k}\left(f^{D}(x_{k})+g^{D}(x_{k},u_{k}^{D})\right).\label{eq:dt_sys_dyn}
\end{eqnarray}
}

\textbf{\textcolor{black}{Assumption 1:}}\textcolor{black}{{} Based
on human ankle kinematic data \cite{winter2009biomechanics}, ankle
position, velocity, and moment are continuous. Therefore, at slow
gait cycle speed, $f^{\zeta}(.)$ and $g^{\zeta}(.)$ are assumed
to be Lipschitz at the switching instant. We utilize this assumption
in subsequent sections to derive the linear predictor model using
Koopman operator. Switching criteria for similar systems with continuous
states, but discrete actuation have been considered in \cite{cousin2019closed},
where the switching between different muscle groups is represented
by their respective actuation matrices which are bounded.}

\section{\textcolor{black}{Koopman-Based Model Predictive Control\protect\label{sec:Koopman-Based-Model-Predictive}}}

\textcolor{black}{This section provides the mathematical framework
for predicting the nonlinear ankle joint dynamics actuated by FES
using Koopman Operator Theory (KOT).}

\subsection{\textcolor{black}{Prediction/Identification}}

\textcolor{black}{We consider the dynamics in \ref{eq:dtsys} where
the controlled state $x_{k}\in\mathcal{X}$, input $u_{k}\in\mathcal{U}$are
sampled to form a finite set, $\mathcal{N}_{c}$. The Koopman operator
acts on a function space $\mathcal{F}$of mapping from $\mathcal{X}$
into $\mathbb{R}$, referred to as observables. The Koopman operator,
$\mathcal{K}$, is an infinite-dimensional linear operator that models
the time-based evolution of a composite function $\varLambda(x_{k})\in\mathbb{R}^{\infty}$,
which act as the $koopman$ $observables$, forward in time. Koopman
operators are parameterized by $x_{k},u_{k}$as follows
\begin{eqnarray}
\mathcal{K}\varLambda(x_{k}) & = & \varLambda(f(x_{k},u_{k})),\qquad\varLambda\in\mathcal{F}\label{Koop_def}
\end{eqnarray}
}

\textcolor{black}{where $\mathcal{K}$ maps observables to the original
state space dynamics in \ref{eq:dtsys}. A subspace is Koopman-invariant
if
\begin{eqnarray}
\mathcal{K}\varLambda & \in & \mathcal{\bar{F}},\quad\forall\varLambda\in\mathcal{\bar{F}},\;\forall u\in\mathcal{U}\label{Koop_inv}
\end{eqnarray}
}

\textcolor{black}{A dictionary of observables, $\chi:\mathcal{X\to\mathbb{R}^{N}}$,
is Koopman-invariant if its elements span a Koopman-invariant subspace.
Since evaluation of the dictionary often involves \textquotedblleft lifting\textquotedblright{}
of the original state vector to a higher dimensional space, $\chi(x_{k})$,
which is commonly referred to as the lifted state. $\chi(x_{k})$
is composed of the original state themselves, or nonlinear functions
of state that are $Lipschitz$ continuous. The extension to non-autonomous
systems has been researched extensively recently, see \cite{proctor2018generalizing,korda2020optimal,mamakoukas2021derivative}.
The extension to non-autonomous system is given as
\begin{eqnarray}
\mathcal{K}(\varLambda(x_{k},u_{k})) & = & \varLambda(f(x_{k},u_{k}),\text{ }h(x_{k},u_{k}))\;\forall\varLambda\in\mathcal{F},\label{Koop_nonaafine}
\end{eqnarray}
}

\textcolor{black}{While this operator renders an infinite-dimensional
system and accurately describes a nonlinear system through a linear
system, but is practically infeasible to implement. For practical
feasibility, the infinite-dimensional operator, $\mathcal{K}$, is
approximated using a finite dimensional operator, defined as $\tilde{\mathcal{K}},$
which is calculated using the Extended Dynamic Mode Decomposition
(EDMD) \cite{proctor2018generalizing}.}

\textcolor{black}{To derive the Koopman operators for each phase,
we collect the time-series data snapshots of the state data as $\{x_{k}\}_{k=1}^{M}$
where $x_{k}$represents the state at time step $k$, and control
input data as $\{u_{k}^{\zeta}\}_{k=1}^{M}$ where $u_{k}^{\zeta}$represents
the control input at time step $k$ during the stance and swing phases.}

\textcolor{black}{We define the lifted-space Koopman observable, $\Psi_{k}(x,u)\in\mathbb{R}^{P}$,
to set up an EDMD problem to predict the linear evolution of the Koopman
observable vector using
\begin{eqnarray}
\Psi_{k+1}(x,u^{\zeta}) & = & \tilde{\mathcal{K}}\Psi_{k}(x,u^{\zeta}),\label{eq:observable_evolve}
\end{eqnarray}
}

\textcolor{black}{where $\tilde{\mathcal{K}}$ is the finite-dimensional
Koopman operator which maps the lifted-state observables forward in
time. Using the state and control time-series snapshots we populate
the lifted-space matrices, as
\begin{eqnarray*}
\mathcal{D^{\zeta}}_{k} & = & \begin{bmatrix}\Psi(x_{1},u_{1}^{\zeta}) & \Psi(x_{2},u_{2}^{\zeta}) & \cdots & \Psi(x_{M-1},u_{M-1}^{\zeta})\end{bmatrix}\\
\mathcal{D}_{k+1}^{\zeta} & = & \begin{bmatrix}\Psi(x_{2},u_{2}^{\zeta}) & \Psi(x_{3},u_{3}) & \cdots & \Psi(x_{M},u_{M})\end{bmatrix}
\end{eqnarray*}
where $\mathcal{D^{\zeta}}_{k}$, $\mathcal{D^{\zeta}}_{k+1}\in\mathbb{R}^{P\times M}$
$\forall\,k=1,\dots,M$, are the collected observable block snapshots
from FES inputs and IMU state measurements for each gait phase. The
Koopman observable vector dynamically evolves as
\begin{equation}
\mathcal{D}_{k+1}=\tilde{\mathcal{K}}\mathcal{D}_{k},\label{eq:koopdef-1}
\end{equation}
where $\mathcal{D}_{k}=\Psi_{k}(x,u)=\begin{bmatrix}\Psi_{x}(x_{k}) & \Psi_{u}(u_{k})\end{bmatrix}^{T}$.
To obtain the control state and control flow maps in the lifted space,
the approximated Koopman operator,$\tilde{\mathcal{K}}$ can be further
subdivided as
\begin{equation}
\tilde{\mathcal{K}}=\begin{bmatrix}\tilde{\mathcal{K}}_{xx} & \tilde{\mathcal{K}}_{xu}\\
\tilde{\mathcal{K}}_{ux} & \tilde{\mathcal{K}}_{uu}
\end{bmatrix},\label{eq:koopman=000020flow}
\end{equation}
}

\textcolor{black}{where $\tilde{\mathcal{K}}_{xx}$ represents the
influence of the state observables, $\Psi_{x}(x_{k}),$ on the future
state observables, and $\tilde{\mathcal{K}}_{xu}$ represents the
influence of the control observables, $\Psi_{u}(u_{k})$, on the future
state observables. The terms $\tilde{\mathcal{K}}_{ux},\tilde{\mathcal{K}}_{uu}$
in \ref{eq:koopman=000020flow} refers to mappings that evolve the
observations on control which are ignored here.}

\textcolor{black}{To determine the Koopman operator for each phase,
$\zeta=\{P-stance\:phase,\,D-swing\:phase\}$, we set up a least-squares
regression problem wherein the error difference between the observed
next step data $\mathcal{D}_{k+1}$, and the prediction from $\tilde{\mathcal{K}}^{\zeta}\mathcal{D}_{k}$,
described as
\begin{eqnarray*}
\tilde{\mathcal{K}}^{\zeta} & = & \arg\min_{\tilde{\mathcal{K}}}\sum_{k=0}^{M-1}\left\Vert \begin{bmatrix}\Psi_{x}(x_{k+1})\\
\Psi_{u}(u_{k+1})
\end{bmatrix}\right.\\
 &  & \left.-\begin{bmatrix}\tilde{\mathcal{K}}_{xx}^{\zeta} & \tilde{\mathcal{K}}_{xu}^{\zeta}\\
\tilde{\mathcal{K}}_{ux}^{\zeta} & \tilde{\mathcal{K}}_{uu}^{\zeta}
\end{bmatrix}\begin{bmatrix}\Psi_{x}(x_{k})\\
\Psi_{u}(u_{k})
\end{bmatrix}\right\Vert ^{2}
\end{eqnarray*}
}

\textcolor{black}{The least-squares solution for $\tilde{\mathcal{K}}$
is given as
\begin{equation}
\tilde{\mathcal{K}}=FG^{\dagger},\label{eq:sol_Koop}
\end{equation}
}

\textcolor{black}{where
\begin{align}
F & =\frac{1}{M}\sum_{k=0}^{M-1}\mathcal{D}_{k+1}\mathcal{D}_{k}^{T},\nonumber \\
G & =\frac{1}{M}\sum_{k=0}^{M-1}\mathcal{D}_{k}\mathcal{D}_{k}^{T},\label{eq:kappa}
\end{align}
}

\textcolor{black}{where pseudoinverse $G^{\dagger}$ is utilized.Using
the indicator function in \ref{eq:indicatorfcn} and the phase-based
Koopman operator, the ankle motion dynamics during a complete gait
cycle can then be represented as
\begin{eqnarray}
\begin{bmatrix}\Psi_{x}(x_{k+1})\end{bmatrix} & = & (1-\sigma_{k})\begin{bmatrix}\tilde{\mathcal{K}}_{xx}^{\text{\ensuremath{P}}} & \tilde{\mathcal{K}}_{xu}^{\ensuremath{P}}\end{bmatrix}\begin{bmatrix}\Psi_{k}(x,u^{P})\end{bmatrix}\nonumber \\
 &  & +\sigma_{k}\begin{bmatrix}\tilde{\mathcal{K}}_{xx}^{\text{\ensuremath{D}}} & \tilde{\mathcal{K}}_{xu}^{\text{\ensuremath{D}}}\end{bmatrix}\begin{bmatrix}\Psi_{k}(x,u^{D})\end{bmatrix}\label{eq:lifted_space_dyn}
\end{eqnarray}
}

\textcolor{black}{To obtain the prediction dynamics for the original
state in (\ref{eq:dtsys}), we compute the flow map between lifted-space
observables,$\Psi_{k}(x,u)$ and original state dynamics, $x_{k}$.
We redefine the state vector $x_{k}$ as $z_{k}$ to avoid any notational
confusion with (\ref{sec:Ankle-joint-dorsiflexion}). To recover $z_{k}$,
we can describe the mapping between Koopman observable, $\Psi_{k}(x,u)$,
and $z_{k}$ as $z_{k}=C\Psi_{k}(x,u)$, where $C\in\mathbb{R}^{3\times P}$denotes
the mapping. To obtain $C$, we solve the following least-squares
problem}

\textcolor{black}{
\begin{align}
\arg\min_{C}\quad\sum_{k=0}^{M-1}\frac{1}{2}\vert\vert C\Psi_{k}(x,u)-z_{k}\vert\vert^{2}.\label{recoverz}
\end{align}
By solving (\ref{recoverz}), and plugging $\Psi_{k}(x,u)=C^{-1}z_{k}$
into the lifted-space flow map \ref{eq:observable_evolve}, we obtain
the linear prediction model for phase-based FES-driven ankle motion
dynamics during a complete gait cycle as
\begin{equation}
z_{k+1}=\tilde{A}^{\zeta}z_{k}+\tilde{B^{\zeta}}u_{k}^{\zeta},\label{eq:phase_linear_predictor}
\end{equation}
where
\begin{eqnarray}
\tilde{A}^{\zeta} & = & \tilde{C\mathcal{K}}_{xx}^{\zeta}C^{-1};\;\tilde{B}^{\zeta}=C\tilde{\mathcal{K}}_{xu}^{\zeta}C^{-1}.\label{eq:linear_maps}
\end{eqnarray}
}

\textcolor{black}{Using the indicator function, $\sigma_{k}$, the
combined state dynamics can be written as
\begin{eqnarray}
z_{k+1} & = & (1-\sigma_{k})\left(\tilde{A}^{P}z_{k}+\tilde{B}^{P}u_{k}^{P}\right)\nonumber \\
 &  & +\sigma_{k}\left(\tilde{A}^{D}z_{k}+\tilde{B}^{D}u_{k}^{D}\right)\label{eq:gen_linear_pred}
\end{eqnarray}
}

\textcolor{black}{where $z_{k}=\left[\begin{array}{cc}
e^{T} & x_{d}^{T}\end{array}\right]^{T}\in\mathbb{R}^{3}$ is the state vector. $A^{P},\,A^{D}\in\mathbb{R}^{3\times3},$}\textbf{\textcolor{black}{$B_{1},\thinspace B_{2}\in\mathbb{R}^{3\times1}$
}}\textcolor{black}{are the Koopman operator based linear state space
mappings, and $u_{k}^{P},\,u_{k}^{D}\in\mathbb{R}$ are the FES control
input vector for assisting ankle plantarflexion and dorsiflexion during
a gait cycle.}

\subsection{\textcolor{black}{Koopman Observables}}

\begin{figure}[t]
\begin{centering}
\textcolor{black}{\includegraphics[scale=0.3]{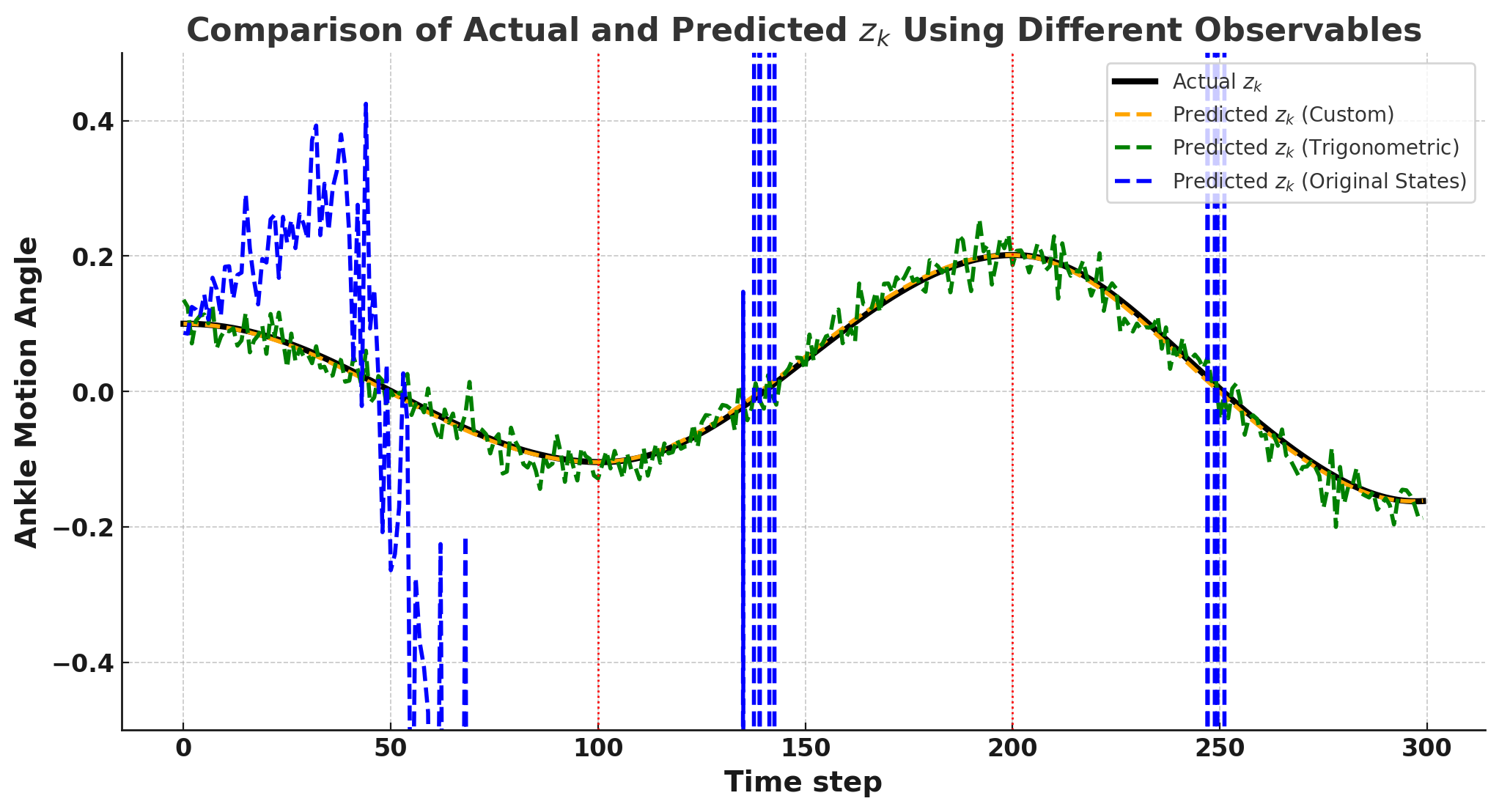}}
\par\end{centering}
\centering{}\textcolor{black}{\caption{\textit{\protect\label{fig:Simulation-results--}Prediction results}
- Plot shows the ankle motion prediction during a gait cycle under
test FES actuation for different observables ($states$, $custom$,
$trigonometric$). The dynamics approximated from (\ref{eq:gen_linear_pred})
are utilized to predict the approximate dynamics.}
}
\end{figure}
\begin{figure}
\begin{centering}
\textcolor{black}{\includegraphics[scale=0.125]{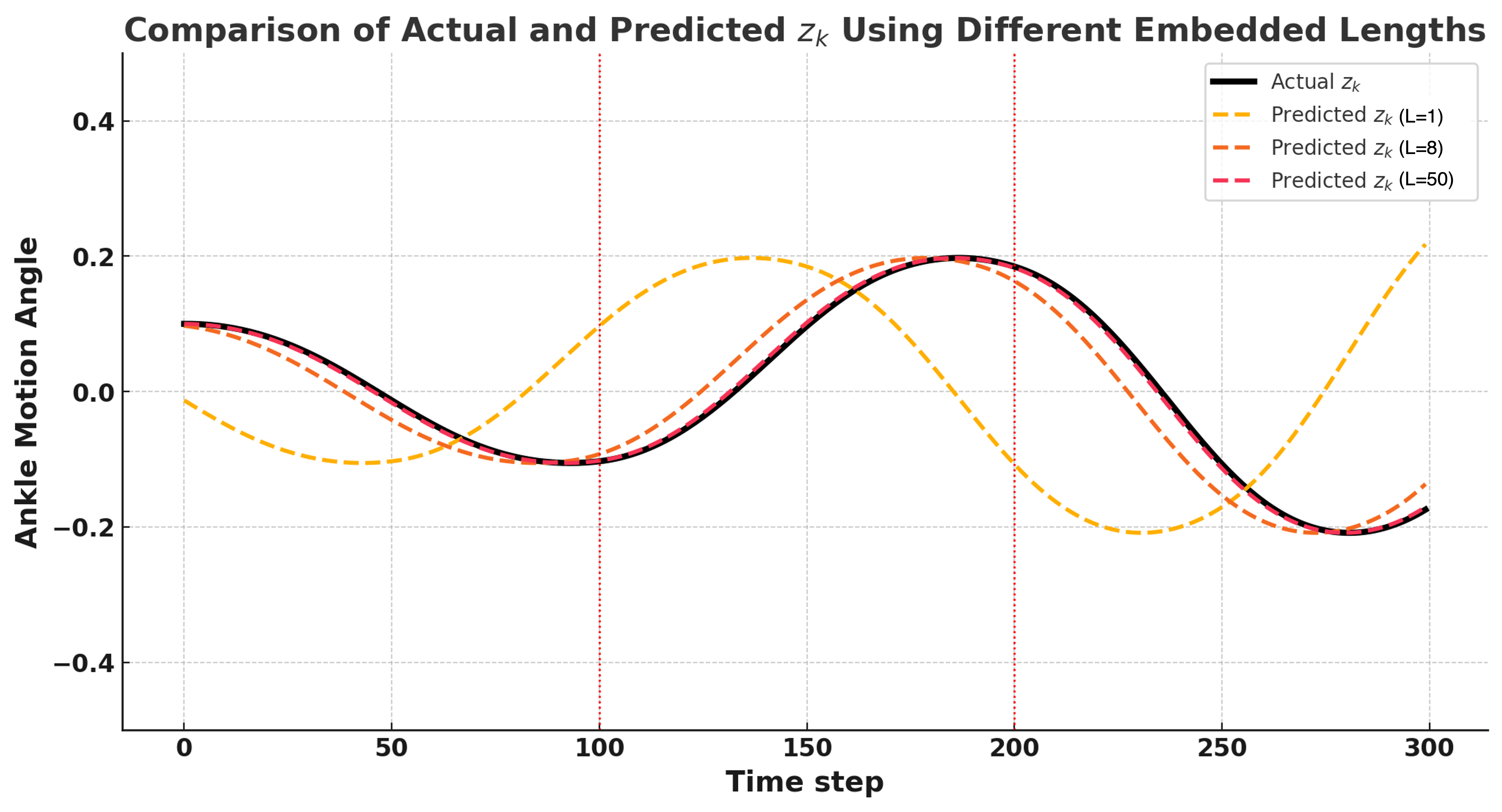}}
\par\end{centering}
\textcolor{black}{\caption{\protect\label{fig:embedding}Comparison of actual and predicted ($z_{k}$)
ankle motion angles using different embedding lengths ($L$) in the
Koopman-based prediction framework. The black solid line represents
the actual ankle motion trajectory, while the dashed lines correspond
to predictions with embedding lengths $L=1$ (yellow), $L=8$ (orange),
and $L=50$ (red). Increasing the embedding length improves prediction
accuracy, as evidenced by the closer alignment of the L = 50 prediction
with the actual trajectory. The results highlight the importance of
appropriate embedding length selection in achieving accurate Koopman-based
predictions.}
}
\end{figure}
\begin{table}[t]
\textcolor{black}{\caption{\protect\label{tab:Summarized-RMSE-mean-1} Root mean square errors
($\vert\vert\tilde{x}_{k}\vert\vert)$ and standard deviation ($\sigma)$for
selecting appropriate dictionary (Dict.). Custom, original state,
Trig. - Trigonometric functions.}
}
\begin{centering}
\begin{tabular}{c|cccccccc}
\hline 
\textcolor{black}{$PF$} & \multicolumn{2}{c}{\textbf{\textcolor{black}{S1}}} & \multicolumn{2}{c}{\textbf{\textcolor{black}{A1}}} & \multicolumn{2}{c}{\textbf{\textcolor{black}{A2}}} & \multicolumn{2}{c}{\textbf{\textcolor{black}{A3}}}\tabularnewline
\hline 
\textbf{\textcolor{black}{Dict.}} & \textcolor{black}{$\vert\vert\tilde{x}_{k}\vert\vert$} & \multicolumn{1}{c|}{\textcolor{black}{$\sigma$}} & \textcolor{black}{$\vert\vert\tilde{x}_{k}\vert\vert$} & \textcolor{black}{$\sigma$} & \textcolor{black}{$\vert\vert\tilde{x}_{k}\vert\vert$} & \textcolor{black}{$\sigma$} & \textcolor{black}{$\vert\vert\tilde{x}_{k}\vert\vert$} & \textcolor{black}{$\sigma$}\tabularnewline
\cline{2-3}
\textcolor{black}{$Custom$} & \textcolor{black}{5.6} & \textcolor{black}{2.1} & \textcolor{black}{6.5} & \textcolor{black}{2.8} & \textcolor{black}{6.2} & \textcolor{black}{2.4} & \textcolor{black}{5.9} & \textcolor{black}{2.1}\tabularnewline
\textcolor{black}{$State$} & \textcolor{black}{14.7} & \textcolor{black}{3.7} & \textcolor{black}{10.8} & \textcolor{black}{3.7} & \textcolor{black}{11.3} & \textcolor{black}{4.2} & \textcolor{black}{9.8} & \textcolor{black}{3.4}\tabularnewline
\cline{9-9}
\textcolor{black}{Trig.} & \textcolor{black}{2.3} & \textcolor{black}{0.8} & \textcolor{black}{1.6} & \textcolor{black}{0.9} & \textcolor{black}{1.8} & \textcolor{black}{0.7} & \textcolor{black}{2.0} & \textcolor{black}{1.1}\tabularnewline
\end{tabular}
\par\end{centering}
\centering{}%
\begin{tabular}{c|cccccccc}
\hline 
\textcolor{black}{$DF$} & \multicolumn{2}{c}{\textbf{\textcolor{black}{S1}}} & \multicolumn{2}{c}{\textbf{\textcolor{black}{A1}}} & \multicolumn{2}{c}{\textbf{\textcolor{black}{A2}}} & \multicolumn{2}{c}{\textbf{\textcolor{black}{A3}}}\tabularnewline
\hline 
\textbf{\textcolor{black}{Dict.}} & \textcolor{black}{$\vert\vert\tilde{x}_{k}\vert\vert$} & \textcolor{black}{$\sigma$} & \textcolor{black}{$\vert\vert\tilde{x}_{k}\vert\vert$} & \textcolor{black}{$\sigma$} & \textcolor{black}{$\vert\vert\tilde{x}_{k}\vert\vert$} & \textcolor{black}{$\sigma$} & \textcolor{black}{$\vert\vert\tilde{x}_{k}\vert\vert$} & \textcolor{black}{$\sigma$}\tabularnewline
\textcolor{black}{$Custom$} & \textcolor{black}{6.9} & \textcolor{black}{2.3} & \textcolor{black}{4.3} & \textcolor{black}{1.8} & \textcolor{black}{5.2} & \textcolor{black}{2.7} & \textcolor{black}{5.7} & \textcolor{black}{2.3}\tabularnewline
\textcolor{black}{$State$} & \textcolor{black}{23.3} & \textcolor{black}{4.9} & \textcolor{black}{17.6} & \textcolor{black}{6.7} & \textcolor{black}{11.3} & \textcolor{black}{5.4} & \textcolor{black}{14.4} & \textcolor{black}{5.1}\tabularnewline
\textcolor{black}{Trig.} & \textcolor{black}{3.4} & \textcolor{black}{1.8} & \textcolor{black}{2.1} & \textcolor{black}{0.8} & \textcolor{black}{2.8} & \textcolor{black}{0.7} & \textcolor{black}{2.5} & \textcolor{black}{1.1}\tabularnewline
\end{tabular}
\end{table}
The choice of basis functions for constructing the dictionary of observables
significantly impacts the performance of the Koopman operator \cite{proctor2018generalizing}.
Appropriate choice for basis functions can be found in \cite{mamakoukas2019local}.
The accuracy of the Koopman operator improves with the length, $P$,
of the observable vector, $\Psi_{k}(x,u)$. As $P\to\infty$, the
Koopman operator, $\tilde{\mathcal{K}}$, accurately describes linear
prediction dynamics for the original nonlinear system \cite{budivsic2012applied}.

\textcolor{black}{We set a prediction accuracy threshold, $\vert\vert x_{k}-z_{k}\vert\vert^{2}\leq\eta$,
for $\eta\leq0.5$ RMSE for ankle motion during a gait cycle. We achieved
the threshold for $P=13$. For the ankle assistance control with state
as joint angles - $\theta(t)$ and $\dot{\theta}(t)$, our Koopman
observable library included a custom library: linear terms - $\theta_{1}$,
$\theta_{2}$, $\dot{\theta}_{1}$,$\dot{\theta}_{2}$ , and nonlinear
terms - $sin(\theta_{1})$, $cos(\theta_{1})$, $sin(\theta_{2})$,
$cos(\theta_{2})$,$\theta_{1}^{2}$, $\theta_{2}^{2}$, $\theta_{1}\theta_{2}$,
$\dot{\theta}_{1}\dot{\theta}_{2}$. The choice of observable is dictated
by the ankle dynamics that exhibit nonlinear effects due to muscle
activation, phase transitions, and joint stiffness. Observables like
$\sin(\theta)$ and $\theta^{2}$ can theoretically capture such effects
by approximating periodic and quadratic relationships seen in ankle
motion, respectively. Including higher-order terms (e.g., $\dot{\theta}^{2}$
, $\sin(\theta)$) helps approximate the nonlinearities associated
with force-length and force-velocity relationships in muscle dynamics.}

\textcolor{black}{Remark: To maintain Koopman invariance, the observables
should be chosen to cover the system\textquoteright s entire dynamic
range, while upholding Assumption 1. Theoretically, the observables
should be designed to provide a stable, controllable Koopman linear
system approximation. Also, to uphold Assumption 1, we use the same
set of observables for both stance and swing phase.}

\subsection{\textcolor{black}{Koopman Model Prediction Accuracy}}

\textcolor{black}{Accuracy of $\mathcal{K}$ is tested with simulation
results. Simulation were performed by using the parameters from \cite{kirsch2017nonlinear}
with different initial conditions to obtain the samples of actual
system trajectories. The dataset used to train the Koopman-based MPC
framework for gait rehabilitation was derived from $150$ gait cycles.
Each gait cycle consists of $200$ samples, with an even split between
the stance and swing phases to capture phase-specific dynamics. This
sampling approach resulted in a dataset with approximately $30,000$
samples. To construct the dataset, we first sample initial states
$(x,\dot{x},u_{fes})\in[25,-20]\times[-2,2]\times[0,50]$. Here, $x$
represents the initial position, $\dot{x}$ represents the initial
velocity, and $u_{fes}$ denotes the FES (Functional Electrical Stimulation)
input level. These ranges were chosen to account for variability in
patient gait patterns, the extent of ankle joint movement during gait
cycles. The control inputs were linearly varied within the range of
$0$ to $30$ $mA$. This ensured that the dataset captured the system\textquoteright s
response to different stimulation levels. A sampling frequency of
$200$ Hz was used. With the generated dataset the Koopman operator
was designed. Based on the prediction dynamics, simulation results
for a nominal sinusoidal trajectory tracking of the ankle joint dynamics
for different observables are given in Fig. (\ref{fig:Simulation-results--}).}

\textcolor{black}{Another important criteria for prediction accuracy
of $\mathcal{\tilde{K}}^{\zeta}$ is the number of past states considered
in the observables referred as the embedding length.We considered
different sample ranges to compare the prediction accuracy of the
approximated Koopman operator, $\mathcal{\tilde{K}}^{\zeta}$, for
different embedding lengths. Prediction for sample ranges $L=1,\,8,\,50$
are plotted in Fig. (\ref{fig:embedding}).}

\section{\textcolor{black}{Data-driven model predictive control\protect\label{sec:Data-driven-model-predictive}}}

\subsection{\textcolor{black}{Koopman Model Predictive Control}}

\textcolor{black}{Let the decision and state variables be defined
as
\begin{eqnarray}
z_{k} & = & [\begin{array}{ccc}
z_{k|k}^{i} & ... & z_{k+N|k}^{i}\end{array}];\label{eq:predvec}\\
u_{k} & = & [\begin{array}{ccc}
u_{k|k}^{i} & ... & u_{k+N-1|k}^{i}\end{array}],
\end{eqnarray}
}

\textcolor{black}{where the vectors $z_{k,}u_{k}\in\mathbb{R}^{3},\:\mathbb{R}$
are the state and control vectors written in the standard MPC notation..
Using the indicator function in (\ref{eq:indicatorfcn}), we can describe
both the stance and swing phase, timed as $[T_{start},t_{stance}]$
and $[t_{swing},T_{end}]$, linear prediction dynamics. The model
predictive problem can then be formulated as follows
\begin{align}
\min_{u^{p},u^{d}}\quad & J(z_{k},u_{k|k})=\sum_{i=1}^{T_{U}}l(.)+\quad V_{T_{N}}\label{eq:ddmpc-1}\\
\text{subject to}\nonumber \\
 & z_{k+1+j|k}=(1-\sigma_{k})\left(A^{P}z_{k}+B^{P}u_{k}^{P}\right)\nonumber \\
 & \qquad+\sigma_{k}\left(A^{D}z_{k}+B^{D}u_{k}^{D}\right) & (a)\\
 & z_{k|k}\in\varOmega_{\chi}^{\zeta},\;u_{k|k}\in\mathcal{\varOmega_{\upsilon}^{\zeta}} & (b)\nonumber \\
 & \Delta z_{k+T_{N}}\in\varOmega_{\chi^{+}}, & (c)\nonumber 
\end{align}
}

\textcolor{black}{where $l(.)=\vert\vert\bar{z}_{k+1}^{T}\vert\vert_{Q_{1}}^{2}+(1-\sigma_{k})\vert\vert u_{k+1}^{P^{T}}\vert\vert_{R^{\zeta}}^{2}+\sigma_{k}\vert\vert u_{k+1}^{D^{T}}\vert\vert_{R^{\zeta}}^{2}$
and $V_{T_{N}}=z_{k+T_{U}}^{T}S^{\zeta}z_{k+T_{U}}$ are the running
and terminal cost. $T_{U}$ is the prediction horizon. Based on the
indicator function, $\sigma_{k}$, $T_{U}$ represents the prediction
horizon for the gait intervals $t_{st}$ and $t_{sw}$. The indicator
function during experiments is implemented based on ground reaction
forces (GRF) which is non-zero during the stance phase and zero during
the swing phase. The running cost, $l(.)$, is the performance measure
penalizing the kinematic state and control inputs considered over
the control horizon, $T_{U}$, for both stance and swing phase. $Q\in\mathbb{R}^{2\times2}$
and $R\in\mathbb{R}$ are }\textit{\textcolor{black}{positive definite}}\textcolor{black}{{}
weighting matrices penalizing the individual states and control inputs
and ensures $l$ and $V$ are positive definite (PD) and radially
unbounded (RU). $S^{\zeta}\in\mathbb{R}^{2\times2}$ is the terminal
cost weighting matrix. $\mathcal{\varOmega_{\upsilon}}$ denotes the
FES stimulation bounds and $\varOmega_{\chi}$ denotes the set of
the state constraints. (As the current time step is fixed based on
the number of samples, $z_{k}$ will be used instead of $z_{k|k}$
, and system matrices derived over $M$ samples will be denoted by
$A^{\zeta},B^{\zeta}$ to simplify the notations). $\varOmega_{\chi^{+}}$
denotes the terminal set defined to ensure that the state remains
within a stabilizable region at the end of the prediction horizon.
Gait phase based terminal weighting matrix
\begin{equation}
S^{\zeta}=\begin{cases}
S^{P}, & \text{if }\sigma_{k}=0\\
S^{D}, & \text{if }\sigma_{k}=1
\end{cases},\label{eq:S_zeta}
\end{equation}
 is derived for both the stance and swing phase by solving the discrete-time
algebraic Riccati equation (DARE)
\begin{eqnarray}
S^{\zeta} & = & \tilde{\mathcal{K}^{T}}_{xx}^{\zeta}S^{\zeta}\tilde{\mathcal{K}}_{xx}^{\zeta}\nonumber \\
 &  & -\tilde{\mathcal{K}^{T}}_{xx}^{\zeta}S^{\zeta}\tilde{\mathcal{K}}_{xu}^{\zeta}(R+\tilde{\mathcal{K}^{T}}_{xu}^{\zeta}S^{\zeta}\tilde{\mathcal{K}}_{xu}^{\zeta})^{-1}\tilde{\mathcal{K}^{T}}_{xu}^{\zeta}S^{\zeta}\tilde{\mathcal{K}}_{x}^{\zeta}\nonumber \\
 &  & +Q,
\end{eqnarray}
where $\tilde{\mathcal{K}}_{xx}^{\zeta},\:\tilde{\mathcal{K}}_{xu}^{\zeta}$
are obtained from the Koopman prediction model}\textit{\textcolor{black}{\emph{.
. We define the terminal set as $\varOmega_{\chi^{+}}=\{z\mid(z_{k+T_{U}})^{T}S^{\zeta}z_{k+T_{U}}\leq\epsilon\},$such
that there exists a stabilizing terminal control law, such as an LQR
policy, where $u_{T_{N}}^{\zeta}=\pi(z_{T_{N-1}})\in\mathcal{\varOmega_{\upsilon}^{\zeta}}$
which ensures that the closed-loop stability criteria, $V_{T_{N}+1}\leq V_{T_{N}}-(l(\bar{z}_{T_{N}},\bar{\tau}_{T_{N}-1},\bar{\phi}_{T_{N}-1}))$,
is satisfied for the MPCA problem in \ref{eq:ddmpc-1}. The set is
parameterized by $\epsilon$ which ensures that the terminal state
remains bounded and controllable.}}}

\section{\textcolor{black}{Experimental Results\protect\label{sec:Experimental-Results}}}

\subsection{\textcolor{black}{Data Collection}}

\begin{figure*}[t]
\begin{centering}
\textcolor{black}{\includegraphics[scale=0.6]{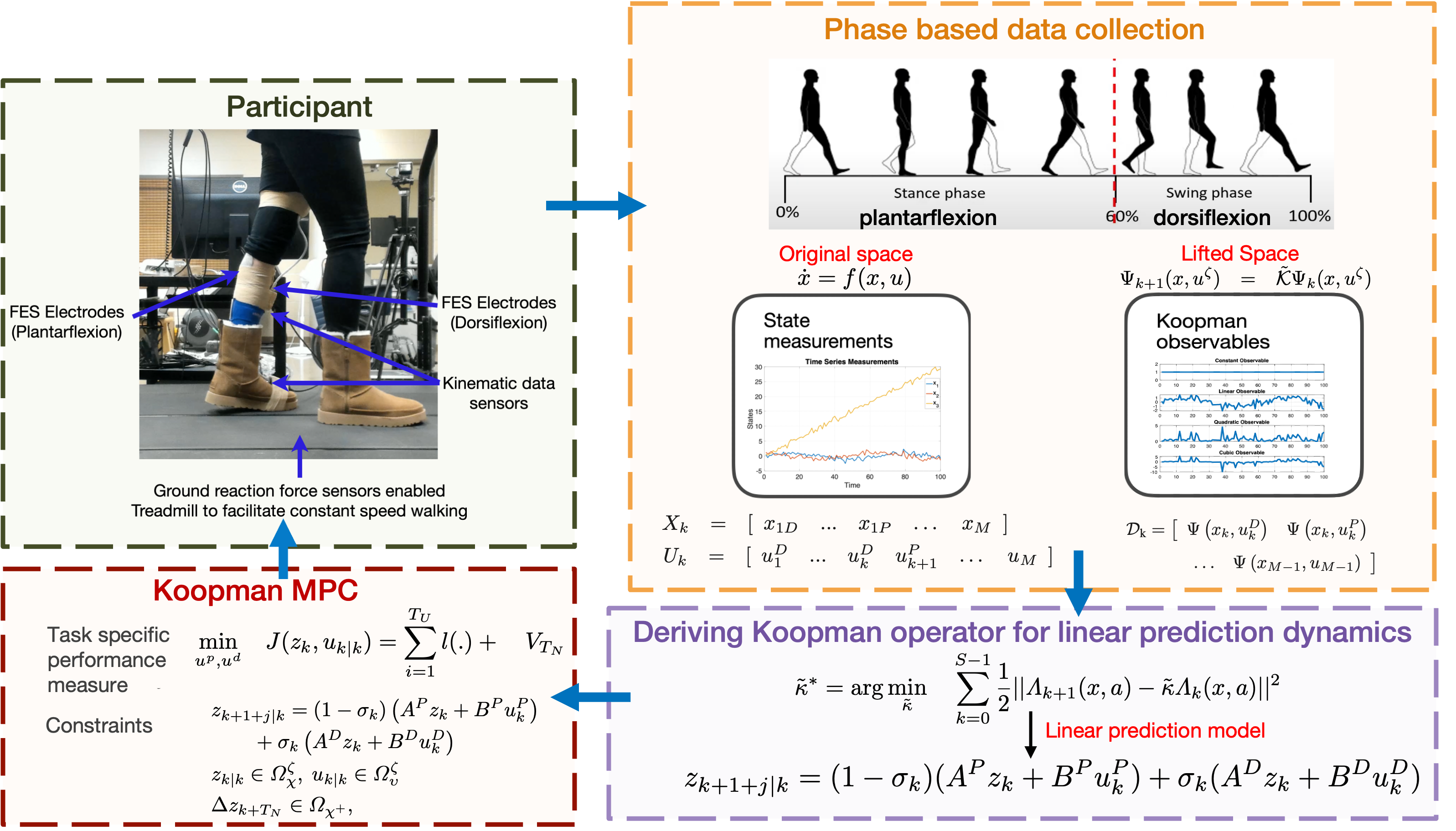}}
\par\end{centering}
\centering{}\textcolor{black}{\caption{\textit{\protect\label{fig:Experimental-task-setup}} \textit{The
experimental setup DDMPC framework for FES-driven gait assistance
are illustrated} - . The participant walks on a treadmill equipped
with ground reaction force (GRF) sensors to detect gait phase transitions
(stance and swing phases). FES electrodes are placed on the Gastrocnemius
(GAS) and Tibialis Anterior (TA) muscles to induce plantarflexion
and dorsiflexion, respectively, with stimulation parameters set at
$f=33$ Hz, $i=u_{k|k}$ mA. Kinematic data sensors record ankle motion
dynamics, while the treadmill enables constant-speed walking. Phase-based
data collection captures state measurements ($x_{k}$) and FES inputs
($u_{k}$) during walking, dividing the gait cycle into stance and
swing phases. The raw data is lifted to a higher-dimensional space
using Koopman observables, which capture nonlinear dynamics in a linear
framework. The Koopman operator predicts the system dynamics via the
lifted representation: $z_{k+1+j|k}=(1-\sigma_{k})(A^{P}z_{k}+B^{P}u_{k}^{P})+\sigma_{k}(A^{D}z_{k}+B^{D}u_{k}^{D})$,
where $\sigma_{k}$ distinguishes stance ($\sigma_{k}=0$) and swing
($\sigma_{k}=1$) phases. The Koopman MPC optimizes FES inputs to
minimize a task-specific performance measure while adhering to state
and control constraints, enabling real-time, phase-specific gait assistance.
This framework effectively coordinates plantarflexion and dorsiflexion
to support natural walking patterns.}
}
\end{figure*}
\textcolor{black}{The study was approved by the Institutional Review
Board (at North Carolina State University (IRB Protocol number: 20602).}

\textsf{\textcolor{black}{$Participants$:}}\textcolor{black}{{} Three
able-bodied subjects (A1, A2, A3, age: 27.4 \textpm{} 3.1 years, height:
1.73 \textpm{} 0.15 m, mass: 82.0 \textpm{} 7.1 kg) without any neuromuscular
or orthopedic disorders were recruited. One subject (S1, age: 62years,
height 1.53 m, mass: 49kg) with multiple sclerosis (MS) was recruited.}

\textsf{\textcolor{black}{$System\:ID\;task$:}}\textcolor{black}{{}
For the data collection pertaining to walking tasks, the experimental
setup was designed to capture the dynamics of gait on a treadmill.
Both able-bodied and MS subjects, designated as A1, A2, and A3, and
S1, respectively, walked at a controlled speed of $0.1\,m/s$, $0.2\,m/s$,
and $0.3\,m/s$ , to accommodate the slow speed and no volition of
subject with Multiple Sclerosis (MS). This setup aimed to collect
comprehensive Inertial Measurement Unit (IMU) data reflecting joint
angles, as well as stimulation currents (plantarflexion = $10-25mA$,
dorsiflexion = $10-20mA$, frequency = $33Hz$) directed at the TA
and GAS muscles. The stimulation parameters, specifically the current
and frequency, were maintained consistently across trials, with FES
stimulation current as decision variable. This approach allowed for
the collection of detailed data on how varying the control input influence
the muscles' response during the walking task for accurate Koopman
operator derivation. Each subject underwent three trials for the first
two sessions to ensure a robust data set for accurate koopman operator
derivation.}

\textcolor{black}{$Experimental\;protocol$: A wearable sensing system,
based on \cite{burner2018wearable}, was used to measure the ankle
joint kinematics. Along with measuring the ankle kinematics, IMU and
ground reaction forces (GRF) measurements were also used for gait
phase detection based on methods discussed in \cite{muller2015experimental},
\cite{seel2016iterative}. A real-time target machine (Speedgoat Inc.,
Liebefeld, Switzerland) was used for experiments with integrated GRF,
IMU signals, and FES stimulation through MATLAB $2019b$. The data
is sampled at a sampling frequency of $200$ Hz. The prediction horizon,
$T_{U}$, is selected based on the average duration of a gait cycle
of individuals, which was an average of $2-4$ seconds for the speeds
$0.1m/s$, $0.2m/s$, and $0.3m/s$. A prediction horizon of $100-200$
ms were chosen.}

\textcolor{black}{To prevent muscle fatigue, particularly in the TA
and GAS muscle, sufficient rest periods were integrated into the experimental
protocol. The treadmill was equipped with GRF sensors. GRF measurements
enabled describing the indicator function, $\sigma_{k}$ which facilitates
precise triggering of the stance and swing phases' FES stimulation.
The data garnered from these walking tasks, including IMU readings
of joint angles and FES stimulation details, were used to populate
the observable matrix.}

\subsection{\textcolor{black}{Experiments \& Results}}

\textcolor{black}{The participants walked on a treadmill with FES
applied on on the TA and GAS muscles during the swing and stance phases,
respectively. The walking setup is illustrated in Fig. (\ref{fig:Experimental-task-setup}).
The FES electrodes were placed on the fibular head and the lateral
malleolus of the TA muscle. For plantarflexion, the negative electrode
is placed on the head of GAS muscle and the positive electrode is
placed above the Achilles tendon. The DDMPC algorithm described in
(\ref{eq:ddmpc-1}) computed FES inputs to TA and GAS muscles. The
switching between them was implemented with ground reaction force
based gait phase detection indicator function, $\sigma_{k}$, to trigger
stance and swing optimal stimulation for plantarflexion and dorsiflexion.
The primary objective of these task was to avoid any foot drag and
achieve adequate foot clearance (pitch, $25deg>x_{1}>-20deg.$) for
each gait cycle during the entire trial.}

\textcolor{black}{We design the desired gait trajectories based on
methods described in \cite{zhang2022ultrasound}. Instead of relying
on a predefined time-dependent trajectory that needs to be adjusted
for each individual or gait cycle, the method uses virtual constraints
based on the orientations and angular velocities of the thigh and
shank segments. The desired trajectory is represented by a Bezier
polynomial, where the parameters of the polynomial are optimized using
a genetic algorithm-based particle swarm optimization (GAPSO) \cite{manyam2021trajectory}.}

\textcolor{black}{The real-time implementation was implemented in
MATLAB/Simulink (R2019b, MathWorks, MA, USA) and executed on a Speedgoat
target machine (Speedgoat Inc., Liebefeld, Switzerland). The Koopman
Model Predictive Control (MPC) was implemented using the Gradient-based
Receding Horizon Model Predictive Control (GRAMPC) solver \cite{englert2019software}.
The GRAMPC algorithm used a prediction horizon of $0.1$ seconds and
a sampling rate of $200$ Hz. The solver uses a gradient-based optimization
approach, dynamically switching between controlling the tibialis anterior
(TA) and gastrocnemius (GAS) muscles based on a ground reaction force
(GRF)-based gait phase detection indicator function.}

\begin{figure*}
\begin{centering}
\textcolor{black}{\includegraphics[scale=0.5]{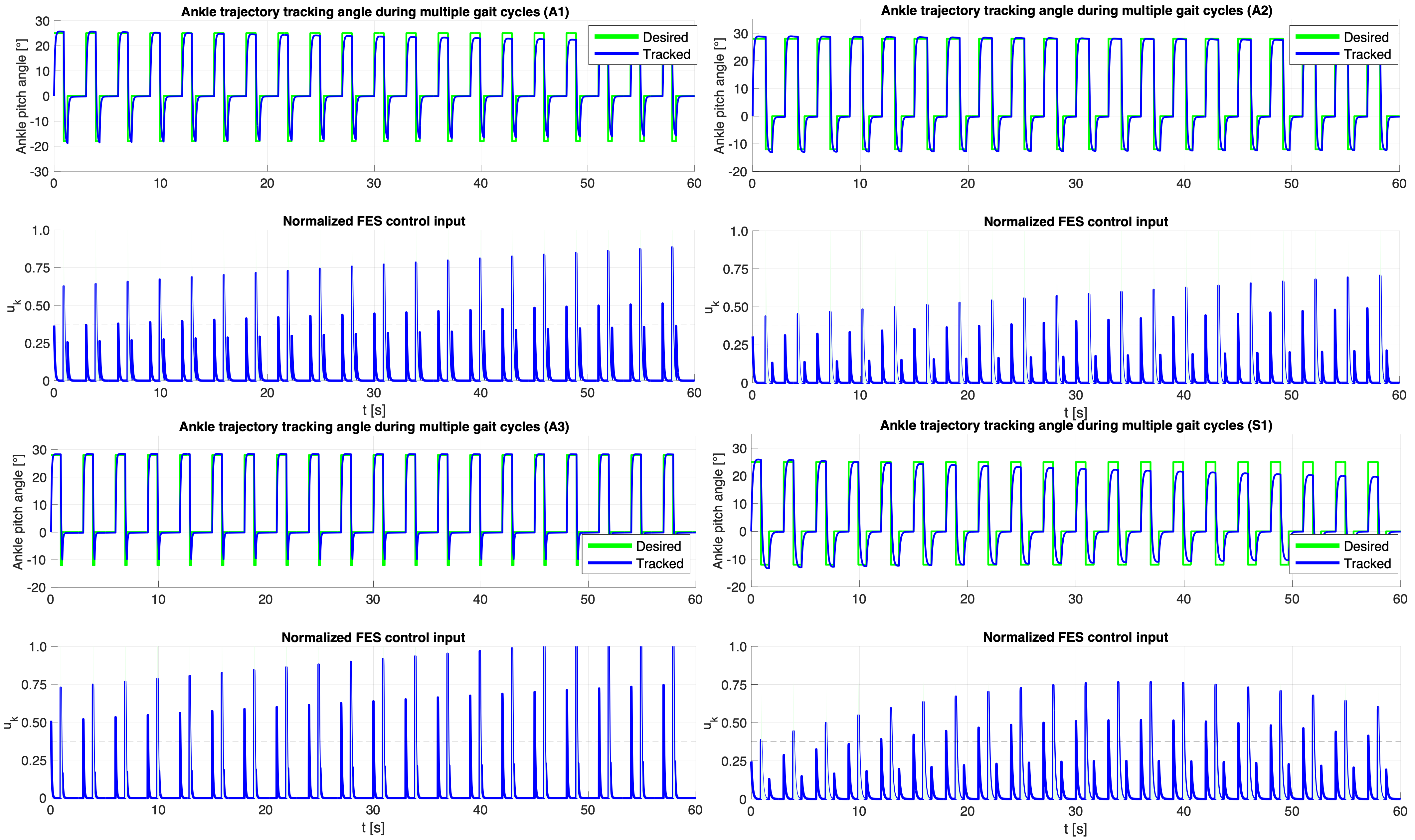}}
\par\end{centering}
\textcolor{black}{\caption{\protect\label{fig:Trajectory-tracking}Trajectory tracking of ankle
motion using DDMPC FES for subjects A1, A2, A3, and S1 (clockwise).
The figure illustrates ankle motion trajectory tracking performance
when the muscles are fully rested. The tracking achieved in this condition
demonstrates a Root Mean Square Error (RMSE) of $1.625^{\circ}$,
highlighting the system\textquoteright s effectiveness in accurate
control under optimal muscle conditions. Note, only absolute values
of FES stimulation, $u$, are provided for both phases.}
}
\end{figure*}

\textcolor{black}{The experiments were divided into $8$ sessions
where the first $2$ sessions were used to generate Koopman operator
characteristics. For implementing DDMPC, in each session we conducted
$4$ trials each at speeds $0.1m/s$, $0.2m/s$, and $0.3m/s$, that
is $12$ trials in total per session. Each trial was conducted with
an interval of 5-7 minutes to recover from muscle fatigue. In total,
$4$ trials each at $3$ different speeds across $3$ sessions were
conducted for each subject, that is, $36$ trials in total. For each
speed the first trial showed the best tracking results. The mean trajectory
tracking plots for both plantarflexion and dorsiflexion for first
trials across all speeds and sessions are presented in Fig. (\ref{fig:Trajectory-tracking})
and the RMSE metrics are presented in (\ref{tab:First=000020trials}).
The treadmill walking speeds in the current study were selected as
$0.1m/s$, $0.2m/s$, and $0.3m/s$, due to the targeted clinical
population with little to no volition in their affected leg. Successive
trials across all sessions and speeds showed a drop in trajectory
tracking due to muscle fatigue.}
\begin{table*}[t]
\textcolor{black}{\caption{\protect\label{tab:First=000020trials}Mean and SD values of ankle
joint trajectory tracking for the mean gait cycle for subjects for
both plantarflexion (left) and dorsiflexion (right) for trial $1$
at speeds $0.1m/s$, $0.2m/s$, and $0.3m/s$ across all sessions.
12 trials per speed in total.(Unit:$\lyxmathsym{\protect\textdegree}$)}
}
\centering{}%
\begin{tabular}{c|cccccc}
\hline 
\textcolor{black}{Speed} & \multicolumn{2}{c}{\textcolor{black}{0.1 m/s}} & \multicolumn{2}{c}{\textcolor{black}{0.2 m/s}} & \multicolumn{2}{c}{\textcolor{black}{0.3 m/s}}\tabularnewline
\hline 
\textcolor{black}{Participant} & \textcolor{black}{Mean} & \textcolor{black}{SD} & \textcolor{black}{Mean} & \textcolor{black}{SD} & \textcolor{black}{Mean} & \textcolor{black}{SD}\tabularnewline
\hline 
\textcolor{black}{S1} & \textcolor{black}{2.3} & \textcolor{black}{0.6} & \textcolor{black}{2.8} & \textcolor{black}{1.4} & \textcolor{black}{6.7} & \textcolor{black}{2.5}\tabularnewline
\hline 
\textcolor{black}{A1} & \textcolor{black}{1.7} & \textcolor{black}{0.7} & \textcolor{black}{2.0} & \textcolor{black}{1.1} & \textcolor{black}{2.8} & \textcolor{black}{1.1}\tabularnewline
\hline 
\textcolor{black}{A2} & \textcolor{black}{1.8} & \textcolor{black}{0.6} & \textcolor{black}{1.9} & \textcolor{black}{0.8} & \textcolor{black}{3.1} & \textcolor{black}{1.4}\tabularnewline
\hline 
\textcolor{black}{A3} & \textcolor{black}{1.5} & \textcolor{black}{0.8} & \textcolor{black}{2.3} & \textcolor{black}{1.2} & \textcolor{black}{3.5} & \textcolor{black}{1.6}\tabularnewline
\hline 
\end{tabular}%
\begin{tabular}{c|cccccc}
\hline 
\textcolor{black}{Speed} & \multicolumn{2}{c}{\textcolor{black}{0.1 m/s}} & \multicolumn{2}{c}{\textcolor{black}{0.2 m/s}} & \multicolumn{2}{c}{\textcolor{black}{0.3 m/s}}\tabularnewline
\hline 
\textcolor{black}{Participant} & \textcolor{black}{Mean} & \textcolor{black}{SD} & \textcolor{black}{Mean} & \textcolor{black}{SD} & \textcolor{black}{Mean} & \textcolor{black}{SD}\tabularnewline
\hline 
\textcolor{black}{S1} & \textcolor{black}{1.4} & \textcolor{black}{0.4} & \textcolor{black}{2.2} & \textcolor{black}{0.8} & \textcolor{black}{4.7} & \textcolor{black}{1.3}\tabularnewline
\hline 
\textcolor{black}{A1} & \textcolor{black}{0.8} & \textcolor{black}{0.2} & \textcolor{black}{1.5} & \textcolor{black}{0.3} & \textcolor{black}{2.5} & \textcolor{black}{0.8}\tabularnewline
\hline 
\textcolor{black}{A2} & \textcolor{black}{0.9} & \textcolor{black}{0.3} & \textcolor{black}{1.9} & \textcolor{black}{0.3} & \textcolor{black}{1.9} & \textcolor{black}{1.1}\tabularnewline
\hline 
\textcolor{black}{A3} & \textcolor{black}{1.1} & \textcolor{black}{0.5} & \textcolor{black}{1.3} & \textcolor{black}{0.4} & \textcolor{black}{2.1} & \textcolor{black}{0.7}\tabularnewline
\hline 
\end{tabular}
\end{table*}
\begin{table}[h]
\textcolor{black}{\caption{\protect\label{tab:Performance=000020trials}Mean and SD values of
ankle joint trajectory tracking for the \textit{Walking Task }for
successive trials (trials 2, 3 and 4) for speeds $0.1m/s$, $0.2m/s$,
and $0.3m/s$. 12 trials per speed per session in total.(Unit:$\lyxmathsym{\protect\textdegree}$)}
}
\centering{}%
\begin{tabular}{c|cccccc}
\hline 
\textcolor{black}{Speed} & \multicolumn{2}{c}{\textcolor{black}{Trial 1 (0.1m/s)}} & \multicolumn{2}{c}{\textcolor{black}{Trial 2(0.1)}} & \multicolumn{2}{c}{\textcolor{black}{Trial 3(0.1)m/s}}\tabularnewline
\hline 
\textcolor{black}{Participant} & \textcolor{black}{Mean} & \textcolor{black}{SD} & \textcolor{black}{Mean} & \textcolor{black}{SD} & \textcolor{black}{Mean} & \textcolor{black}{SD}\tabularnewline
\hline 
\textcolor{black}{S1} & \textcolor{black}{4.6} & \textcolor{black}{1.2} & \textcolor{black}{5.2} & \textcolor{black}{1.6} & \textcolor{black}{11.2} & \textcolor{black}{2.8}\tabularnewline
\hline 
\textcolor{black}{A1} & \textcolor{black}{2.3} & \textcolor{black}{1.3} & \textcolor{black}{2.8} & \textcolor{black}{1.6} & \textcolor{black}{2.6} & \textcolor{black}{1.5}\tabularnewline
\textcolor{black}{A2} & \textcolor{black}{2.8} & \textcolor{black}{1.3} & \textcolor{black}{3.5} & \textcolor{black}{1.3} & \textcolor{black}{4.7} & \textcolor{black}{2.1}\tabularnewline
\textcolor{black}{A3} & \textcolor{black}{3.2} & \textcolor{black}{1.7} & \textcolor{black}{4.2} & \textcolor{black}{2.4} & \textcolor{black}{4.5} & \textcolor{black}{2.6}\tabularnewline
\hline 
\end{tabular}
\end{table}
\begin{figure*}
\begin{centering}
\textcolor{black}{\includegraphics[scale=0.27]{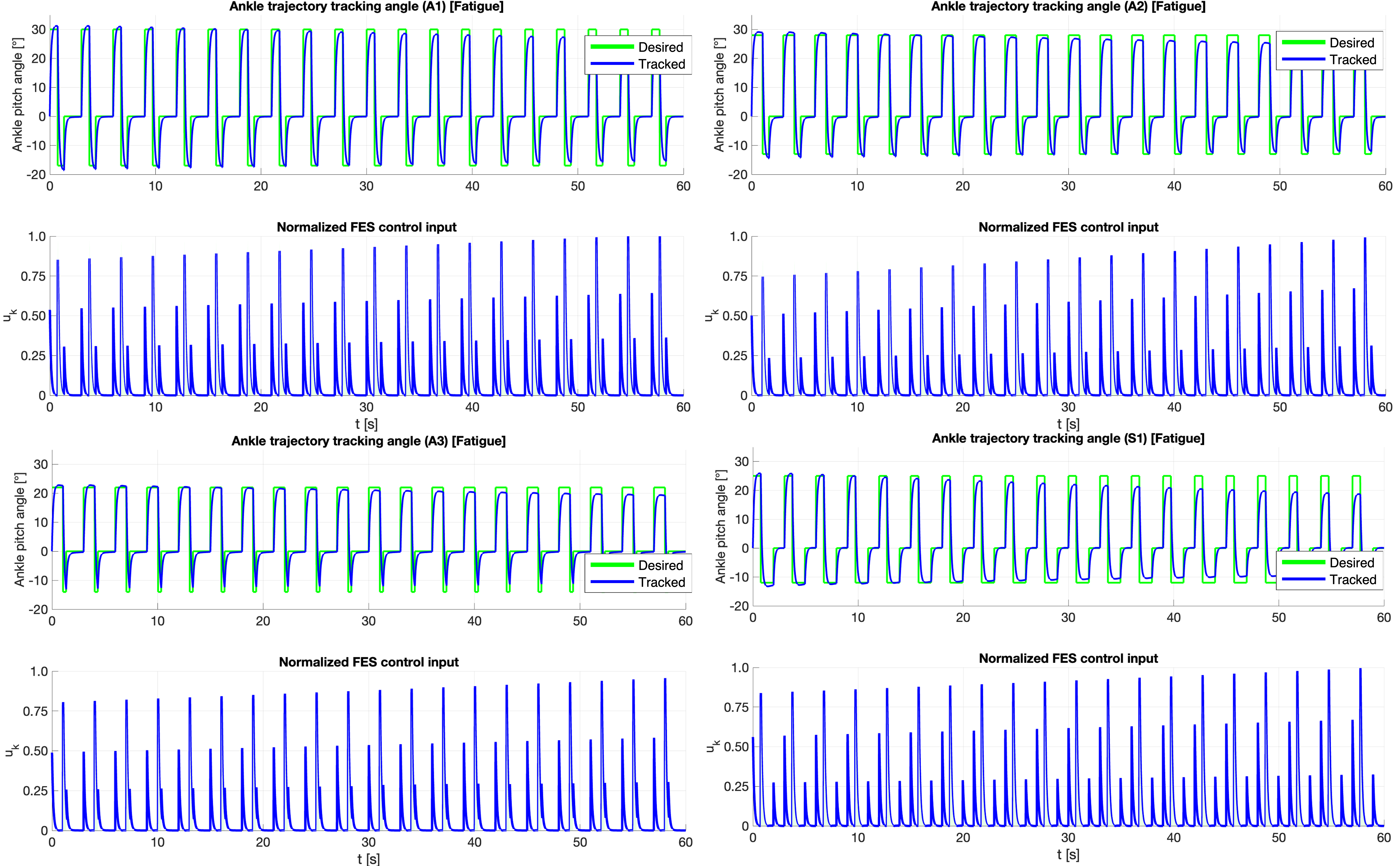}}
\par\end{centering}
\textcolor{black}{\caption{\protect\label{fig:fatiguetrack}Ankle motion trajectory tracking
results averaged over the final trial of each session for each participant
A1, A2, A3, and S1 (\textit{clockwise}).. The figure illustrates trajectory
tracking performance after $3\lyxmathsym{\textendash}4$ walking trials
of 60 seconds each, reflecting the effects of muscle fatigue. The
trajectory Root Mean Square Error (RMSE) is $3.1^{\circ}$, indicating
the onset of fatigue-induced deviations in tracking accuracy.}
}
\end{figure*}
\begin{figure*}
\begin{centering}
\textcolor{black}{\includegraphics[scale=0.22]{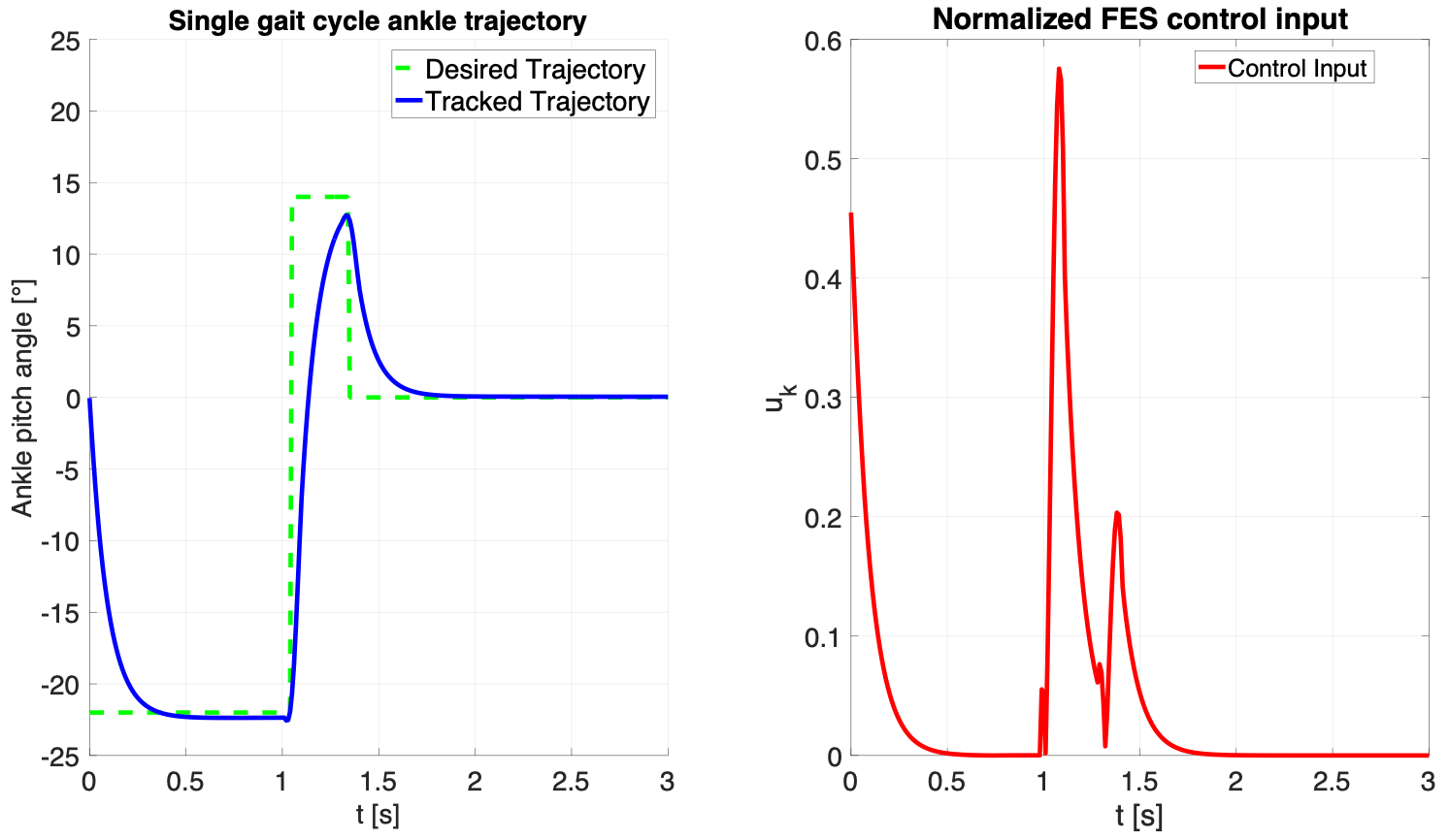}}
\par\end{centering}
\textcolor{black}{\caption{\protect\label{fig:singlegaitcycle}Comparison of the desired and
tracked ankle pitch angle trajectories during a single gait cycle,
achieved using a Koopman-based Model Predictive Control (MPC) framework.}
}
\end{figure*}
\begin{figure*}
\begin{centering}
\textcolor{black}{\includegraphics[scale=0.25]{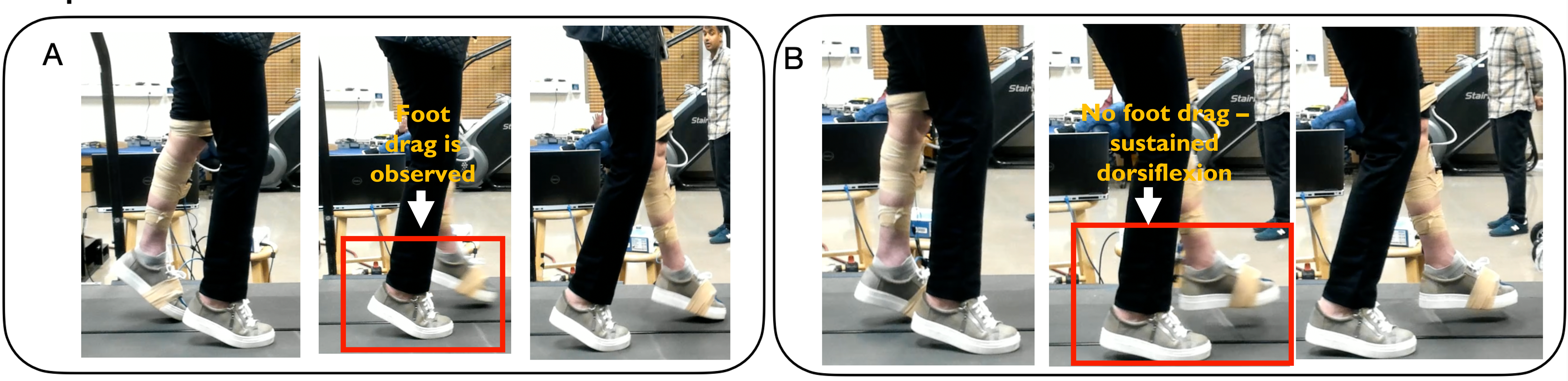}}
\par\end{centering}
\textcolor{black}{\caption{\protect\label{fig:pred_post}Comparison of gait performance for Subject
S1 before and after using the FES-driven gait assist controller. \textbf{A}:
Foot/toe drag observed prior to FES application, where the subject
was unable to sustain walking on a treadmill at the lowest speed of
$0.1\,\mathrm{ms}^{-1}$. B: No foot/toe drag observed after applying
the FES-driven controller, enabling sustained walking at treadmill
speeds of $0.1,\,0.2,\,0.3\,\mathrm{ms}^{-1}$. Annotations highlight
the transition between pre-FES and post-FES conditions and the effectiveness
of the proposed controller in supporting walking performance.}
}
\end{figure*}
\textcolor{black}{Trajectory tracking showed consistent ankle plantarflexion
and dorsiflexion response actuated by FES using Koopman MPC. We observe
that FES input saturated only for participant A3 but maintained good
trajectory tracking. FES input for TA muscle always remained within
the prescribed limits, which is an improvement to our past results
presented in \cite{singh2023data} and shows the benefit of using
gait-specific MPC controller to design FES input ankle assistance
during gait constrain the inputs. Effect of FES-driven gait assist
in S1 is described in Fig. (\ref{fig:pred_post}).}

\textcolor{black}{Fig. (\ref{fig:singlegaitcycle}) shows the trajectory
tracking results for a single gait cycle. The reference trajectory
consists of set points representing the adequate plantarflexion and
dorsiflexion angles for the stance, swing, and rest phases of the
gait cycle (green dashed line). The tracked trajectory (blue solid
line) demonstrates the controller\textquoteright s ability to accurately
follow these set points. Subplot (}\textit{\textcolor{black}{left}}\textcolor{black}{)
illustrates the trajectory tracking performance, while subplot (}\textit{\textcolor{black}{right}}\textcolor{black}{)
shows the corresponding control input (red solid line) applied to
achieve the tracking.Discussion\label{sec:Discussion}In this work,
we used the KOT approach that can efficiently linearize the nonlinear
dynamics of human ankle allowing for the application of a linear MPC
strategy for both plantarflexion and dorsiflexion control. This linearization
facilitates the formulation of the MPC problem as a real-time solvable
quadratic program. This approach also offers a high degree of adaptability.
By continuously incorporating new data, the model can dynamically
adjust to changes in the patient's gait, such as variations in walking
speed. This makes the system highly personalized, as it can cater
to the specific requirements and progress of each individual patient.
This approach is particularly suited to the complex neuromuscular
ankle motion dynamics as it accounts for human variability in muscle
response due to FES stimulations, but doesn't actually require the
exact individual system parameters.}

\textcolor{black}{We hypothesize that incorporating volitional muscle
activity should lead to optimal design of GAS and TA FES stimulation
levels which mitigate muscle fatigue effects which is a future direction
for this work. For S1, we observe that swing phase is consistently
of longer duration as compared to able bodied subjects. Moreover,
the trajectory tracking performance deteriorated over time. This is
as expected as there is no volition for S1 in their left ankle. We
now intend to combine a closed-loop ultrasound informed muscle activity
information, described in \cite{zhang2022ultrasound}, in our data-driven
optimal FES control framework to improve trajectory tracking for longer
duration of walking and at higher speed.}

\subsection{\textcolor{black}{Limitations and Future Work}}

\textcolor{black}{The Koopman operator framework provides a linear
prediction model for nonlinear dynamical systems, enabling effective
integration with Model Predictive Control (MPC). However, it faces
certain challenges, such as the need for finite-dimensional approximations
of an inherently infinite-dimensional operator. The appropriate selection
of observables that accurately capture the system\textquoteright s
dynamics remains critical for achieving robust and precise predictions.
Limitations also arise in handling muscle fatigue and real-time variability
in neuromuscular behavior, particularly in dynamic and repetitive
tasks like gait rehabilitation.}

\textcolor{black}{Future research will aim to enhance the Koopman
MPC framework by integrating real-time feedback from physiological
sensors, such as surface electromyography (sEMG) and ultrasound, to
account for muscle activation and fatigue dynamics. Developing adaptive
Koopman operator update laws that incorporate this physiological data
will improve the MPC controller\textquoteright s robustness and adaptability
to changing muscle conditions. Incorporating muscle fatigue models
directly as observables or leveraging real-time sensor feedback will
enable dynamic adjustments to stimulation strategies, mitigating fatigue
effects during repetitive gait cycles.}

\textcolor{black}{Additionally, addressing stability challenges introduced
by switched dynamics between stance and swing phases---especially
at faster gait speeds---will require the development of phase-specific
stability laws based on minimum dwell time based Lyapunov methods.
Future directions also include extending the Koopman MPC framework
for multi-joint control and exploring scalable solutions for higher
degrees of freedom.}

\section{\textcolor{black}{Conclusion\protect\label{sec:Conclusion}}}

\textcolor{black}{We developed a data-driven Model Predictive Control
(MPC) framework to assist with achieving the normal range of ankle
motion during gait. Our approach leverages Koopman Operator Theory
(KOT) to transform the inherently complex and nonlinear dynamics of
FES-actuated ankle motion into a linearized representation. This linearization
enables the application of efficient linear control techniques to
a highly nonlinear system. The linear prediction model derived through
KOT allowed us to formulate the MPC problem as a quadratic program,
significantly enhancing the real-time feasibility, precision, and
adaptability of the FES-driven control system.}

\textcolor{black}{The effectiveness and stability of our approach
were validated through experimental trials involving three participants
without disabilities and one participant with multiple sclerosis (MS).
The results demonstrated precise trajectory tracking assistance for
the developed Koopman MPC controller. The developed KOT-based MPC
framework can be used to deliver effective, real-time, and personalized
assistance for individuals with gait-related impairments, including
those caused by MS, stroke, and incomplete spinal cord injury (SCI).}

\textcolor{black}{\bibliographystyle{IEEEtran}
\bibliography{TNSRE}

\begin{thebibliography}{10}
\providecommand{\url}[1]{#1}
\csname url@samestyle\endcsname
\providecommand{\newblock}{\relax}
\providecommand{\bibinfo}[2]{#2}
\providecommand{\BIBentrySTDinterwordspacing}{\spaceskip=0pt\relax}
\providecommand{\BIBentryALTinterwordstretchfactor}{4}
\providecommand{\BIBentryALTinterwordspacing}{\spaceskip=\fontdimen2\font plus
\BIBentryALTinterwordstretchfactor\fontdimen3\font minus
  \fontdimen4\font\relax}
\providecommand{\BIBforeignlanguage}[2]{{%
\expandafter\ifx\csname l@#1\endcsname\relax
\typeout{** WARNING: IEEEtran.bst: No hyphenation pattern has been}%
\typeout{** loaded for the language `#1'. Using the pattern for}%
\typeout{** the default language instead.}%
\else
\language=\csname l@#1\endcsname
\fi
#2}}
\providecommand{\BIBdecl}{\relax}
\BIBdecl

\bibitem{york2019survey}
G.~York and S.~Chakrabarty, ``A survey on foot drop and functional electrical
  stimulation,'' \emph{International Journal of Intelligent Robotics and
  Applications}, vol.~3, no.~1, pp. 4--10, 2019.

\bibitem{hayashibe2011dual}
M.~Hayashibe, Q.~Zhang, and C.~Azevedo-Coste, ``Dual predictive control of
  electrically stimulated muscle using biofeedback for drop foot correction,''
  in \emph{2011 IEEE/RSJ International Conference on Intelligent Robots and
  Systems}.\hskip 1em plus 0.5em minus 0.4em\relax IEEE, 2011, pp. 1731--1736.

\bibitem{kesar2009functional}
T.~M. Kesar, R.~Perumal, D.~S. Reisman, A.~Jancosko, K.~S. Rudolph, J.~S.
  Higginson, and S.~A. Binder-Macleod, ``Functional electrical stimulation of
  ankle plantarflexor and dorsiflexor muscles: effects on poststroke gait,''
  \emph{Stroke}, vol.~40, no.~12, pp. 3821--3827, 2009.

\bibitem{luo2020review}
S.~Luo, H.~Xu, Y.~Zuo, X.~Liu, and A.~H. All, ``A review of functional
  electrical stimulation treatment in spinal cord injury,''
  \emph{Neuromolecular medicine}, vol.~22, pp. 447--463, 2020.

\bibitem{gil2020advances}
J.~Gil-Castillo, F.~Alnajjar, A.~Koutsou, D.~Torricelli, and J.~C. Moreno,
  ``Advances in neuroprosthetic management of foot drop: a review,''
  \emph{Journal of neuroengineering and rehabilitation}, vol.~17, pp. 1--19,
  2020.

\bibitem{seel2016iterative}
T.~Seel, C.~Werner, J.~Raisch, and T.~Schauer, ``Iterative learning control of
  a drop foot neuroprosthesis generating physiological foot motion in paretic
  gait by automatic feedback control,'' \emph{Control Engineering Practice},
  vol.~48, pp. 87--97, 2016.

\bibitem{page2020point}
A.~Page and C.~Freeman, ``Point-to-point repetitive control of functional
  electrical stimulation for drop-foot,'' \emph{Control Engineering Practice},
  vol.~96, p. 104280, 2020.

\bibitem{jiang2020iterative}
C.~Jiang, M.~Zheng, Y.~Li, X.~Wang, L.~Li, and R.~Song, ``Iterative adjustment
  of stimulation timing and intensity during fes-assisted treadmill walking for
  patients after stroke,'' \emph{IEEE Transactions on Neural Systems and
  Rehabilitation Engineering}, vol.~28, no.~6, pp. 1292--1298, 2020.

\bibitem{muller2020adaptive}
P.~M{\"u}ller, A.~J. Del~Ama, J.~C. Moreno, and T.~Schauer, ``Adaptive
  multichannel fes neuroprosthesis with learning control and automatic gait
  assessment,'' \emph{Journal of neuroengineering and rehabilitation}, vol.~17,
  pp. 1--20, 2020.

\bibitem{zhang2022ultrasound}
Q.~Zhang, K.~Lambeth, A.~Iyer, Z.~Sun, and N.~Sharma, ``Ultrasound
  imaging-based closed-loop control of functional electrical stimulation for
  drop foot correction,'' \emph{IEEE Transactions on Control Systems
  Technology}, 2022.

\bibitem{awad2020central}
L.~N. Awad, H.~Hsiao, and S.~A. Binder-Macleod, ``Central drive to the paretic
  ankle plantarflexors affects the relationship between propulsion and walking
  speed after stroke,'' \emph{Journal of neurologic Physical Therapy}, vol.~44,
  no.~1, pp. 42--48, 2020.

\bibitem{bajd1994significance}
T.~Bajd, A.~Kralj, T.~Kar{\v{c}}nik, R.~{\v{S}}avrin, and P.~Obreza,
  ``Significance of fes-assisted plantarflexion during walking of incomplete
  sci subjects,'' \emph{Gait \& Posture}, vol.~2, no.~1, pp. 5--10, 1994.

\bibitem{proctor2018generalizing}
J.~L. Proctor, S.~L. Brunton, and J.~N. Kutz, ``Generalizing koopman theory to
  allow for inputs and control,'' \emph{SIAM Journal on Applied Dynamical
  Systems}, vol.~17, no.~1, pp. 909--930, 2018.

\bibitem{korda2020optimal}
M.~Korda and I.~Mezi{\'c}, ``Optimal construction of koopman eigenfunctions for
  prediction and control,'' \emph{IEEE Transactions on Automatic Control},
  vol.~65, no.~12, pp. 5114--5129, 2020.

\bibitem{korda2018linear}
M.~Korda and I.~Mezic, ``Linear predictors for nonlinear dynamical systems:
  Koopman operator meets model predictive control,'' \emph{Automatica},
  vol.~93, pp. 149--160, 2018.

\bibitem{benoussaad2013nonlinear}
M.~Benoussaad, K.~Mombaur, and C.~Azevedo-Coste, ``Nonlinear model predictive
  control of joint ankle by electrical stimulation for drop foot correction,''
  in \emph{2013 IEEE/RSJ International Conference on Intelligent Robots and
  Systems}.\hskip 1em plus 0.5em minus 0.4em\relax IEEE, 2013, pp. 983--989.

\bibitem{singh2023data}
M.~Singh and N.~Sharma, ``Data-driven model predictive control for drop foot
  correction,'' in \emph{2023 American Control Conference (ACC)}.\hskip 1em
  plus 0.5em minus 0.4em\relax IEEE, 2023, pp. 2615--2620.

\bibitem{kirsch2017nonlinear}
N.~Kirsch, N.~Alibeji, and N.~Sharma, ``Nonlinear model predictive control of
  functional electrical stimulation,'' \emph{Control Engineering Practice},
  vol.~58, pp. 319--331, 2017.

\bibitem{winter2009biomechanics}
D.~A. Winter, \emph{Biomechanics and motor control of human movement}.\hskip
  1em plus 0.5em minus 0.4em\relax John wiley \& sons, 2009.

\bibitem{cousin2019closed}
C.~A. Cousin, V.~H. Duenas, C.~A. Rouse, M.~J. Bellman, P.~Freeborn, E.~J. Fox,
  and W.~E. Dixon, ``Closed-loop cadence and instantaneous power control on a
  motorized functional electrical stimulation cycle,'' \emph{IEEE Transactions
  on Control Systems Technology}, vol.~28, no.~6, pp. 2276--2291, 2019.

\bibitem{mamakoukas2021derivative}
G.~Mamakoukas, M.~L. Castano, X.~Tan, and T.~D. Murphey, ``Derivative-based
  koopman operators for real-time control of robotic systems,'' \emph{IEEE
  Transactions on Robotics}, vol.~37, no.~6, pp. 2173--2192, 2021.

\bibitem{mamakoukas2019local}
G.~Mamakoukas, M.~Castano, X.~Tan, and T.~Murphey, ``Local koopman operators
  for data-driven control of robotic systems,'' in \emph{Robotics: science and
  systems}, 2019.

\bibitem{budivsic2012applied}
M.~Budisic, R.~Mohr, and I.~Mezic, ``Applied koopmanism,'' \emph{Chaos: An
  Interdisciplinary Journal of Nonlinear Science}, vol.~22, no.~4, p. 047510,
  2012.

\bibitem{burner2018wearable}
L.~Burner and N.~Sharma, ``A wearable sensing system to estimate lower limb
  state for drop foot correction,'' \emph{Highlighting Undergraduate Research
  at the University of Pittsburgh Swanson School of Engineering}, p.~16, 2018.

\bibitem{muller2015experimental}
P.~Muller, T.~Steel, and T.~Schauer, ``Experimental evaluation of a novel
  inertial sensor based realtime gait phase detection algorithm,'' in
  \emph{Proceedings of the Technically Assisted Rehabilitation Conference},
  2015.

\bibitem{manyam2021trajectory}
S.~G. Manyam, D.~W. Casbeer, I.~E. Weintraub, and C.~Taylor, ``Trajectory
  optimization for rendezvous planning using quadratic b{\'e}zier curves,'' in
  \emph{2021 IEEE/RSJ International Conference on Intelligent Robots and
  Systems (IROS)}.\hskip 1em plus 0.5em minus 0.4em\relax IEEE, 2021, pp.
  1405--1412.

\bibitem{englert2019software}
T.~Englert, A.~V{\"o}lz, F.~Mesmer, S.~Rhein, and K.~Graichen, ``A software
  framework for embedded nonlinear model predictive control using a
  gradient-based augmented lagrangian approach (grampc),'' \emph{Optimization
  and Engineering}, vol.~20, pp. 769--809, 2019.

\end{thebibliography}
}
\end{document}